\documentclass[twocolumn,times]{aastex63}
\usepackage{amsmath}
\usepackage{makecell}
\usepackage{multirow,booktabs}

\setwatermarkfontsize{1.5in}

\newcommand{\insight}{\textit{Insight}-HXMT}

\shortauthors{Zheng ET AL.}
\begin{document}

\title{Observation of GRB 221009A early afterglow in X/$\gamma$-ray energy band}

\correspondingauthor{Shao-Lin Xiong, Cheng-Kui Li}
\email{xiongsl@ihep.ac.cn, lick@ihep.ac.cn}

\author{Chao Zheng}
\affil{Key Laboratory of Particle Astrophysics, Institute of High Energy Physics, Chinese Academy of Sciences, Beijing 100049, China}
\affil{University of Chinese Academy of Sciences, Chinese Academy of Sciences, Beijing 100049, China}

\author{Yan-Qiu Zhang}
\affil{Key Laboratory of Particle Astrophysics, Institute of High Energy Physics, Chinese Academy of Sciences, Beijing 100049, China}
\affil{University of Chinese Academy of Sciences, Chinese Academy of Sciences, Beijing 100049, China}

\author{Shao-Lin Xiong*}
\affil{Key Laboratory of Particle Astrophysics, Institute of High Energy Physics, Chinese Academy of Sciences, Beijing 100049, China}

\author{Cheng-Kui Li*}
\affil{Key Laboratory of Particle Astrophysics, Institute of High Energy Physics, Chinese Academy of Sciences, Beijing 100049, China}

\author{He Gao}
\affil{Department of Astronomy, Beijing Normal University, Beijing 100875, People’s Republic of China}

\author{Wang-Chen Xue}
\affil{Key Laboratory of Particle Astrophysics, Institute of High Energy Physics, Chinese Academy of Sciences, Beijing 100049, China}
\affil{University of Chinese Academy of Sciences, Chinese Academy of Sciences, Beijing 100049, China}

\author{Jia-Cong Liu}
\affil{Key Laboratory of Particle Astrophysics, Institute of High Energy Physics, Chinese Academy of Sciences, Beijing 100049, China}
\affil{University of Chinese Academy of Sciences, Chinese Academy of Sciences, Beijing 100049, China}

\author{Chen-Wei Wang}
\affil{Key Laboratory of Particle Astrophysics, Institute of High Energy Physics, Chinese Academy of Sciences, Beijing 100049, China}
\affil{University of Chinese Academy of Sciences, Chinese Academy of Sciences, Beijing 100049, China}

\author{Wen-Jun Tan}
\affil{Key Laboratory of Particle Astrophysics, Institute of High Energy Physics, Chinese Academy of Sciences, Beijing 100049, China}
\affil{University of Chinese Academy of Sciences, Chinese Academy of Sciences, Beijing 100049, China}

\author{Wen-Xi Peng}
\affil{Key Laboratory of Particle Astrophysics, Institute of High Energy Physics, Chinese Academy of Sciences, Beijing 100049, China}

\author{Zheng-Hua An}
\affil{Key Laboratory of Particle Astrophysics, Institute of High Energy Physics, Chinese Academy of Sciences, Beijing 100049, China}

\author{Ce Cai}
\affil{College of Physics and Hebei Key Laboratory of Photophysics Research and Application, Hebei Normal University, Shijiazhuang, Hebei 050024, China}

\author{Ming-Yu Ge}
\affil{Key Laboratory of Particle Astrophysics, Institute of High Energy Physics, Chinese Academy of Sciences, Beijing 100049, China}

\author{Dong-Ya Guo}
\affil{Key Laboratory of Particle Astrophysics, Institute of High Energy Physics, Chinese Academy of Sciences, Beijing 100049, China}

\author{Yue Huang}
\affil{Key Laboratory of Particle Astrophysics, Institute of High Energy Physics, Chinese Academy of Sciences, Beijing 100049, China}

\author{Bing Li}
\affil{Key Laboratory of Particle Astrophysics, Institute of High Energy Physics, Chinese Academy of Sciences, Beijing 100049, China}

\author{Ti-Pei Li}
\affil{Key Laboratory of Particle Astrophysics, Institute of High Energy Physics, Chinese Academy of Sciences, Beijing 100049, China}
\affil{Department of Astronomy, Tsinghua University, Beijing 100084, People’s Republic of China}

\author{Xiao-Bo Li}
\affil{Key Laboratory of Particle Astrophysics, Institute of High Energy Physics, Chinese Academy of Sciences, Beijing 100049, China}

\author{Xin-Qiao Li}
\affil{Key Laboratory of Particle Astrophysics, Institute of High Energy Physics, Chinese Academy of Sciences, Beijing 100049, China}
 
\author{Xu-Fang Li}
\affil{Key Laboratory of Particle Astrophysics, Institute of High Energy Physics, Chinese Academy of Sciences, Beijing 100049, China}

\author{Jin-Yuan Liao}
\affil{Key Laboratory of Particle Astrophysics, Institute of High Energy Physics, Chinese Academy of Sciences, Beijing 100049, China}

\author{Cong-Zhan Liu}
\affil{Key Laboratory of Particle Astrophysics, Institute of High Energy Physics, Chinese Academy of Sciences, Beijing 100049, China}

\author{Fang-Jun Lu}
\affil{Key Laboratory of Particle Astrophysics, Institute of High Energy Physics, Chinese Academy of Sciences, Beijing 100049, China}

\author{Xiang Ma}
\affil{Key Laboratory of Particle Astrophysics, Institute of High Energy Physics, Chinese Academy of Sciences, Beijing 100049, China}

\author{Rui Qiao}
\affil{Key Laboratory of Particle Astrophysics, Institute of High Energy Physics, Chinese Academy of Sciences, Beijing 100049, China}

\author{Li-Ming Song}
\affil{Key Laboratory of Particle Astrophysics, Institute of High Energy Physics, Chinese Academy of Sciences, Beijing 100049, China}

\author{Jin Wang}
\affil{Key Laboratory of Particle Astrophysics, Institute of High Energy Physics, Chinese Academy of Sciences, Beijing 100049, China}

\author{Ping Wang}
\affil{Key Laboratory of Particle Astrophysics, Institute of High Energy Physics, Chinese Academy of Sciences, Beijing 100049, China}

\author{Xi-Lu Wang}
\affil{Key Laboratory of Particle Astrophysics, Institute of High Energy Physics, Chinese Academy of Sciences, Beijing 100049, China}

\author{Yue Wang}
\affil{Key Laboratory of Particle Astrophysics, Institute of High Energy Physics, Chinese Academy of Sciences, Beijing 100049, China}
\affil{University of Chinese Academy of Sciences, Chinese Academy of Sciences, Beijing 100049, China}

\author{Xiang-Yang Wen}
\affil{Key Laboratory of Particle Astrophysics, Institute of High Energy Physics, Chinese Academy of Sciences, Beijing 100049, China}

\author{Shuo Xiao}
\affil{Guizhou Provincial Key Laboratory of Radio Astronomy and Data Processing, 
Guizhou Normal University, Guiyang 550001, People’s Republic of China}
\affil{School of Physics and Electronic Science, Guizhou Normal University, Guiyang 550001, People’s Republic of China}

\author{Yan-Bing Xu}
\affil{Key Laboratory of Particle Astrophysics, Institute of High Energy Physics, Chinese Academy of Sciences, Beijing 100049, China}

\author{Yu-Peng Xu}
\affil{Key Laboratory of Particle Astrophysics, Institute of High Energy Physics, Chinese Academy of Sciences, Beijing 100049, China}

\author{Zhi-Guo Yao}
\affil{Key Laboratory of Particle Astrophysics, Institute of High Energy Physics, Chinese Academy of Sciences, Beijing 100049, China}

\author{Qi-Bing Yi}
\affil{Key Laboratory of Particle Astrophysics, Institute of High Energy Physics, Chinese Academy of Sciences, Beijing 100049, China}
\affil{Key Laboratory of Stellar and Interstellar Physics and Department of Physics, Xiangtan University, Xiangtan 411105, China}

\author{Shu-Xu Yi}
\affil{Key Laboratory of Particle Astrophysics, Institute of High Energy Physics, Chinese Academy of Sciences, Beijing 100049, China}

\author{Yuan You}
\affil{Key Laboratory of Particle Astrophysics, Institute of High Energy Physics, Chinese Academy of Sciences, Beijing 100049, China}

\author{Fan Zhang}
\affil{Key Laboratory of Particle Astrophysics, Institute of High Energy Physics, Chinese Academy of Sciences, Beijing 100049, China}

\author{Jin-Peng Zhang}
\affil{Department of Engineering Physics, Tsinghua University, Beijing 100084, People’s Republic of China}

\author{Peng Zhang}
\affil{Key Laboratory of Particle Astrophysics, Institute of High Energy Physics, Chinese Academy of Sciences, Beijing 100049, China}
\affil{College of Electronic and Information Engineering, Tongji University, Shanghai 201804, China}

\author{Shu Zhang}
\affil{Key Laboratory of Particle Astrophysics, Institute of High Energy Physics, Chinese Academy of Sciences, Beijing 100049, China}
 
\author{Shuang-Nan Zhang}
\affil{Key Laboratory of Particle Astrophysics, Institute of High Energy Physics, Chinese Academy of Sciences, Beijing 100049, China}

\author{Yan-Ting Zhang}
\affil{Key Laboratory of Particle Astrophysics, Institute of High Energy Physics, Chinese Academy of Sciences, Beijing 100049, China}
\affil{University of Chinese Academy of Sciences, Chinese Academy of Sciences, Beijing 100049, China}

\author{Zhen Zhang}
\affil{Key Laboratory of Particle Astrophysics, Institute of High Energy Physics, Chinese Academy of Sciences, Beijing 100049, China}

\author{Xiao-Yun Zhao}
\affil{Key Laboratory of Particle Astrophysics, Institute of High Energy Physics, Chinese Academy of Sciences, Beijing 100049, China}

\author{Yi Zhao}
\affil{Key Laboratory of Particle Astrophysics, Institute of High Energy Physics, Chinese Academy of Sciences, Beijing 100049, China}
\affil{Department of Astronomy, Beijing Normal University, Beijing 100875, China}

\author{Shi-Jie Zheng}
\affil{Key Laboratory of Particle Astrophysics, Institute of High Energy Physics, Chinese Academy of Sciences, Beijing 100049, China}

\begin{abstract} 
The early afterglow of a Gamma-ray burst (GRB) can provide critical information on the jet and progenitor of the GRB. The extreme brightness of GRB 221009A allows us to probe its early afterglow in unprecedented detail. In this letter, we report comprehensive observation results of the early afterglow of GRB 221009A (from $T_0$+660 s to $T_0$+1860 s, where $T_0$ is the \textit{Insight}-HXMT/HE trigger time) in X/$\gamma$-ray energy band (from 20 keV to 20 MeV) by \textit{Insight}-HXMT/HE, GECAM-C and \textit{Fermi}/GBM. We find that the spectrum of the early afterglow in 20 keV-20 MeV could be well described by a cutoff power-law with an extra power-law which dominate the low and high energy bands respectively. The cutoff power-law $E_{\rm peak}$ is $\sim$ 30 keV and the power-law photon index is $\sim$ 1.8 throughout the early afterglow phase. By fitting the light curves in different energy bands, we find that a significant achromatic break (from 10s keV to TeV) is required at $T_0$ + 1246$^{+27}_{-26}$ s (i.e. 1021 s since the afterglow starting time $T_{\rm AG}$=$T_0$+225 s), providing compelling evidence of a jet break. Interestingly, both the pre-break and post-break decay slopes vary with energy, and these two slopes become closer in the lower energy band, making the break less identifiable. Intriguingly, the spectrum of the early afterglow experienced a slight hardening before the break and a softening after the break. These results provide new insights into the physics of this remarkable GRB.

\end{abstract}


\section{Introduction} \label{sec:intro}
Gamma-ray bursts (GRBs), as the cosmological electromagnetic events produced by stellar deaths, are the most violent explosions in the universe \cite[e.g.][]{2018pgrb.book.....Z}. They can be simply divided into short GRBs (SGRBs) and long GRBs (LGRBs) by their duration of prompt emission, which is generally produced by ultra-relativistic and collimated outflow (jet) launched from binary neutron star mergers or massive star core collapses, respectively \citep{1989Nucleosynthesis, 1992Gamma, 1993Gamma, 2006ARA&A..44..507W, 2017ApJ...848L..13A}. While the jet propagates away from the central engine, the prompt emission is produced within the jet by internal shock or magnetic dissipation \citep{Rees:1994nw, zhang2011} and multi-wavelength afterglow is radiated when the jet sweeps the external medium, producing reverse shock and forward shock \cite[][for a review]{gao2013NewAR}. 

In contrast to the irregular behavior in the prompt emission, the afterglow emission usually can be well described with several power-law decays. Sometimes the break between power-law decays is achromatic (i.e. showing up in different energy bands at the same time), called a jet break, which is directly related to the jet geometry and observer's viewing angle \citep{Panaitescu:2001bx, Liang:2007rn, 2018Gamma}. Generally, when the jet opening angle is small, the jet break occurs earlier in the light curve, making the multi-wavelength observation of the break very challenging.

As the brightest GRB event since the discovery of GRBs in the 1960s, GRB 221009A was first reported by \textit{Fermi}/GBM while many other gamma-ray telescopes detected it simultaneously but without real-time alert. With the early accurate location provided by $Swift$/BAT \citep{GCN.32632}, follow-up observations have been made by numerous multi-wavelength and multi-messenger telescopes \cite[e.g.][]{HXMT-GECAM, malesani2023brightest, williams2023grb, Burns:2023oxn, science.adg9328, Kann:2023ulv, Levan:2023doz, Negro:2023cer, Laskar:2023yap}. The redshift of this burst is found to be $z$=0.151 \citep{2022GCN.32648....1D, 2023arXiv230207891M}.

The \textit{Insight}-Hard X-ray Modulation Telescope (\insight) was triggered by GRB 221009A at 13:17:00.050 UT on October 9th, 2022 (denoted as $T_0$\footnote{Note that the \textit{Insight}-HXMT trigger is 60 ms after the \textit{Fermi}/GBM trigger time for GRB 221009A.}) during a routine ground search of bursts. With the novel design dedicated to extremely bright bursts, GECAM-C made a uniquely accurate measurement of GRB 221009A prompt emission with its unsaturated data and found that this GRB has the highest total isotropic-equivalent energy ($E_\gamma,_{\rm iso}$=1.5 × 10$^{55}$ erg) \citep{HXMT-GECAM}, which is confirmed by other observations (e.g. \cite{Burns:2023oxn, GBM_09A, Frederiks:2023bxg, Rodi:2023jzx}). Interestingly, a jet break at $T_{\rm AG}$ + 670$_{-110}^{+230}$ s is reported in TeV range \citep{science.adg9328}, where $T_{\rm AG}$ is the afterglow starting time\footnote{The accurate afterglow starting time $T_{\rm AG}$ = $T_0$ + 225.7 s was reported by LHAASO \citep{science.adg9328}. We use $T_{\rm AG}$ = $T_0$ + 225 s in this work. The slightly different choice of $T_{\rm AG}$ has negligible impact on our results.}. Joint observations of early afterglow by \insight~and GECAM-C also show there is a jet break around $T_{\rm AG}$ + 950$_{-50}^{+60}$ s in the keV-MeV range, however, a detailed analysis of the jet break is difficult due to observation gaps of these two instruments \citep{HXMT-GECAM}.

In this paper, we performed a comprehensive analysis of the GRB 221009A early afterglow (from $T_0$+660 s to $T_0$+1860 s
%, where $T_0$ is trigger time
) in X-ray and gamma-ray energy band (20 keV to 20 MeV) with joint observations of the \textit{Insight}-HXMT High Energy X-ray Telescope, GECAM-C Telescope, and the Gamma-ray Burst Monitor of Fermi Gamma-ray Space Telescope. We present the detailed observations of the three instruments and the data analysis in section \ref{section3}. Then temporal and spectral analysis results are reported in section \ref{section4}. Discussion and conclusions are given in section \ref{section5}. 

Please note that all parameter errors in this work are for 68\% confidence level if not otherwise stated.

\section{Observations and data reduction} \label{section3}
\subsection{\textit{Insight}-HXMT/HE observation}
As China's first X-ray astronomy satellite, \textit{Insight}-Hard X-ray Modulation Telescope (\textit{Insight}-HXMT) has made many discoveries on high energy objects since it was launched to the orbit of an altitude of 550 km and an inclination of 43 degrees on June 15th, 2017  \citep{HXMT_mission_zhang, HXMT_overview_zhang, HE_calibration_Li, 2023In}. The main detector of the High Energy X-ray telescope (HE), consisting of 18 NaI(Tl)/CsI(Na) phoswich scintillation detectors, has played an important role in the observation of GRB 221009A in the MeV range \citep{HXMT-GECAM}.

GRB 221009A was visible to HE before the satellite entered the Earth's shadow. However, the effective observation of GRB 221009A afterglow is up to about $T_0$+900 s, when \insight~ began to enter the high-latitude region where the HE detectors suffer high background. The high fluence of charged particles in the high-latitude orbit region had a significant impact on observation data \citep{HXMT-GECAM}. The incident angle of GRB 221009A to HE detector is shown in Fig. \ref{Mutipl_sat_angle} panel $(a)$ during the early afterglow phase, when \textit{Insight}-HXMT was executing a galactic plane scan observation.

Excluding the time region of background contamination, HE data from $T_0$+660 s to $T_0$+900 s are selected for this analysis, as shown in Fig. \ref{Mutipl_sat_lc} panel $(a)$. The blue solid lines represent the afterglow light curves in the 600-3000 keV energy band, and the corresponding background (black lines) is estimated by a parametric model, which makes use of some measured parameters to reconstruct the background \citep{HXMT_bkg, 2023arXiv230214459L}. For the spectral analysis, the HE data in the good time interval (GTI) is split into two time-resolved spectra as shown in Table \ref{afterglow_table}, which are used for joint analysis with \textit{Fermi}/GBM data.

\subsection{\textit{Fermi}/GBM observation}
As one of the two instruments onboard the Fermi Gamma-ray Space Telescope, the Gamma-ray Burst Monitor (\textit{Fermi}/GBM) is composed of 14 detectors with different orientations: 12 Sodium Iodide (NaI) detectors (labeled from N0 to N11) covering the energy range of about 8-1000 keV, and 2 Bismuth Germanate (BGO) detectors (labeled as B0 and B1) covering energies about 0.2-40 MeV \citep{Fermi/GBM_overview, Fermi/GBM_calibration}. \textit{Fermi}/GBM was also triggered by GRB 221009A, with the observation results initially reported on the GCN \citep{GCN.32636} and detailed analysis presented in \cite{GBM_09A}. 

Since GRB 221009A is extremely bright, GBM suffered pulse pile-up and data saturation during the prompt emission and flare emission phase \citep{GBM_09A}. Fortunately, the GBM observation of the early afterglow was unaffected by these effects and thus could provide valuable data until it entered the Earth's shadow at about $T_0$+1467 s. 

The GBM Time-Tagged Events (TTE) data are performed for temporal and spectral analysis, with two NaI detectors (NaI 4, NaI 8) and two BGO detectors (BGO 0, BGO 1). The incident angles of these four detectors are shown with the red solid lines in Fig. \ref{Mutipl_sat_angle} panel $(b)$ to $(e)$ respectively. The green and blue dotted lines indicate the nearest revisit orbit geographically (with a time shift of $\pm$85610 s w.r.t the current orbit of the detection of GRB 221009A), showing that the detector pointing directions were nearly identical. Since the environmental irradiation background was also stable, the count rate in the revisit orbit can be used to estimate the background of the observation of GRB 221009A. This method of background estimation has been proven to be successful for GECAM-C \citep{HXMT-GECAM}. Indeed it is also validated by the GBM light curve itself before $T_0$ and after the Earth occultation (see Fig. \ref{Mutipl_sat_lc} panel $(b)$ to $(e)$).

The light curves from $T_0$-300 s to $T_0$+2000 s for N4, N8, B0, and B1 are plotted in Fig. \ref{Mutipl_sat_lc} on panel $(b)$ to panel $(e)$. The gray solid lines represent the total light curve and the black solid lines represent the background light curve averaged from the revisit orbits. As shown in red lines, we also tried a background fit with the same 4th-order polynomial with the same background time intervals as Figure 3 of \cite{GBM_09A}. Because the 4th-order polynomial function is unable to accurately describe the background after the flare, our averaged revisit orbit background seems to be more appropriate to describe the afterglow background. Considering the influence of the Sco X-1 and charged particle events, we omit data below 200 keV for NaI 4. The NaI 8 data below 20 keV is also omitted \cite[]{GBM_09A}. BGO data below 400 keV are also ignored since the detector response at low energies is affected by the photomultiplier tubes and their housings at such viewing angle \cite[]{GBM_09A}. For the spectral analysis, three time-resolved spectra, as shown in Table \ref{afterglow_table}, are extracted from $T_0$+660 s to $T_0$+1400 s with $GBMDataTools$ \footnote{ \href{https://fermi.gsfc.nasa.gov/ssc/data/analysis/gbm/}{https://fermi.gsfc.nasa.gov/ssc/data/analysis/gbm/}} \citep{GbmDataTools}. Additionally, finer time-resolved spectra with a time interval of 40 s are achieved with $GBMDataTools$ from $T_0$+660 s to $T_0$+1420 s, which are used to characterize the spectra evolution in 20-300 keV energy band.

\subsection{GECAM-C observation}
As the third member of the GECAM constellation, GECAM-C\footnote{GECAM-C is also named as HEBS \citep{GCN.32751}.} onboard the SATech-01 satellite was launched on July 27th, 2022. GECAM-C is equipped with 12 Gamma-Ray detectors (GRDs, labeled from GRD01 to GRD12, including 6 LaBr$_3$-based detectors and 6 NaI-based detectors) and 2 Charged Particle Detectors (CPDs)  \citep{zhangdl:2023GECAMC,zheng:2023ground_calibration, Zhangyq:2023cross_calibration}. 

\begin{figure}[http]
\centering
\includegraphics[width=\columnwidth]{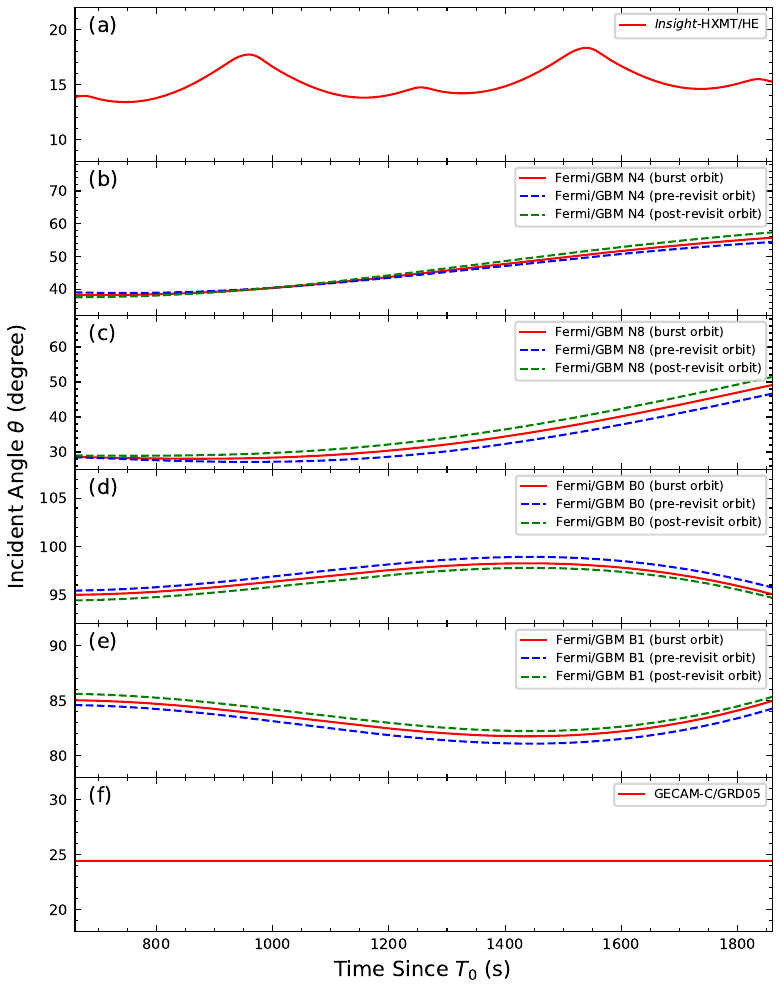}
\caption{During the observations of GRB 221009A afterglow, incident angles of \insight, \textit{Fermi}/GBM N4, N8, \textit{Fermi}/GBM B0, B1, and GECAM-C GRD05 are shown with red lines. The dotted lines with green and blue represent the incident angle of \textit{Fermi}/GBM N4, N8, B0, and B1 in the pre-revisit orbit and post-revisit orbit respectively.}
\label{Mutipl_sat_angle}
\end{figure}

\begin{figure*}
\centering
\begin{minipage}[t]{0.48\textwidth}
\centering
\includegraphics[width=\columnwidth]{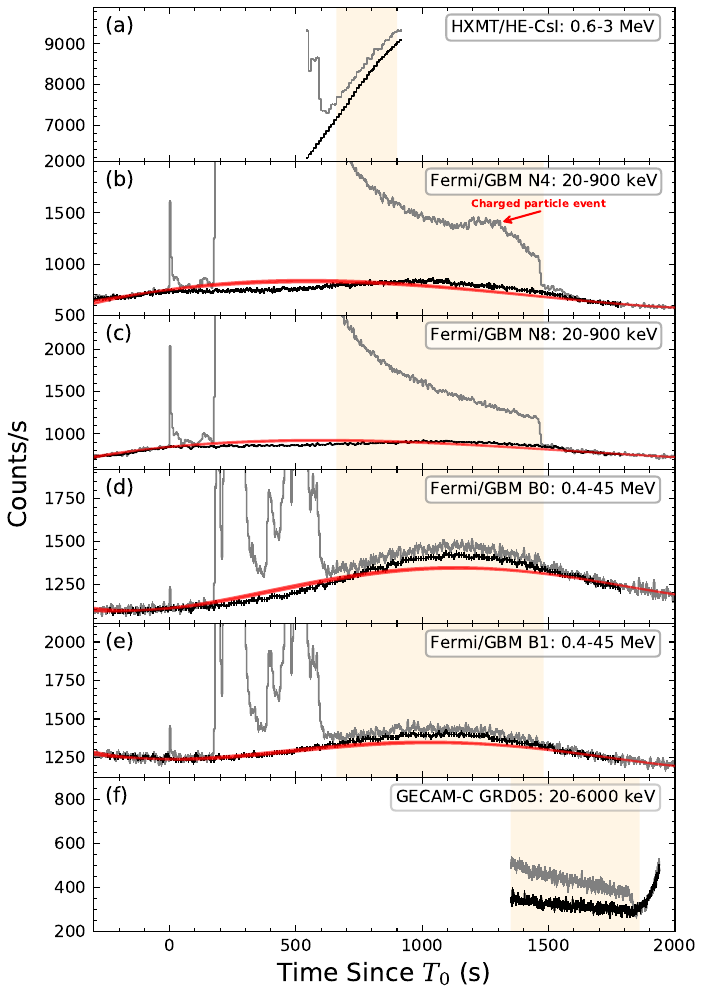}
\end{minipage}
\begin{minipage}[t]{0.48\textwidth}
\centering
\includegraphics[width=\columnwidth]{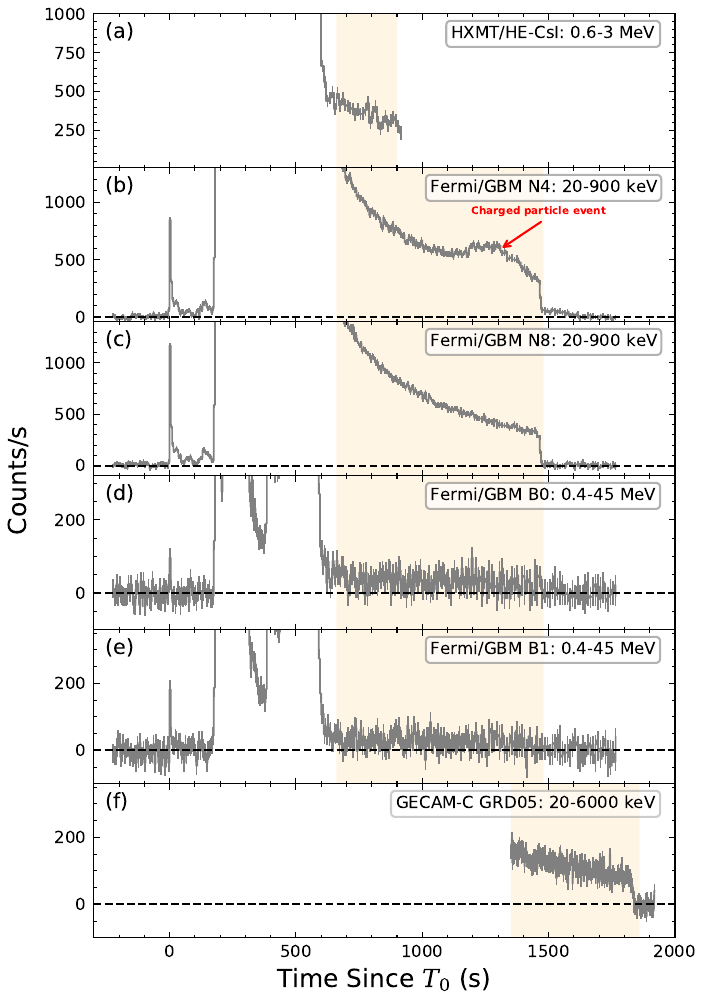}
\end{minipage}
\caption{\textbf{Left}: The gray lines in all panels are total light curves of \insight/HE (0.6-3 MeV), Fermi/GBM N4 (20-900 keV), N8 (20-900 keV), B0 (0.4-45 MeV), B1 (0.4-45 MeV) and GECAM-C GRD05 (20-6000 keV) during the observation of GRB 221009A. The black light curves are the background of the revisit orbit after the trigger time when the satellites were almost in the same orbital locations and orientations. The red lines show the background fitted with the 4th-order polynomial. The orange shaded area in each panel marks the time windows which are confirmed as the early afterglow and is used for spectral analysis. \textbf{Right}: The net light curves of GRB 221009A, for which the background is estimated from the averaged revisit orbital background.}
\label{Mutipl_sat_lc}
\end{figure*}

Thanks to the dedicated design in instrument and working mode for extremely bright burst observation, the GECAM-C made an accurate measurement of GRB 221009A and provided unique data in the main burst, flare, and early afterglow until it entered the Earth's shadow at about $T_{\rm 0}$+1860 s \citep{GCN.32751}. From $T_0$+640 s to $T_0$+1350 s, the GECAM-C was over the high latitude region \citep{HXMT-GECAM}, where the flux of charged particle events was extremely high and contaminated some of the observation data \citep{HXMT-GECAM}. Consequently, we select clean data of GRD05 from $T_0$+1350 s to $T_0$+1860 s for temporal and spectral analysis. The incident angle of GRD05 was stable during this time range as shown in Fig. \ref{Mutipl_sat_angle} on panel $(d)$. The light curves of GRD05 are displayed in Fig. \ref{Mutipl_sat_lc} on panel $(g)$, which gray lines are the total light curves and black solid lines are the background light curves using revisit orbit data \citep{wangcw:2023bkg, HXMT-GECAM}. The total light curves agree well with the background light curves after GECAM-C entered the Earth's shadow, supporting that the revisit orbit background is appropriate.

\section{Results} \label{section4}
\subsection{Spectral analysis}
The observation data are divided into four parts, and the time-resolved spectra are obtained (see Table.\ref{afterglow_table}). For the spectral analysis, the single power-law (Eq.\ref{equ:pl_Model}) model (PL hereafter) and the cutoff power-law (Eq.\ref{equ:Cutoffpl_Model}) model (CPL hereafter), and Band model (Eq.\ref{equ:band_Model}) is performed to describe the spectral features:

%powerlaw model
\begin{equation}   
A(E)=KE^{-\alpha},
\label{equ:pl_Model}
\end{equation}
where $K$ is the normalization amplitude constant ($\rm photons \cdot cm^{-2} \cdot s^{-1} \cdot keV^{-1}$) at 1 keV, $\alpha$ is the dimensionless photon index of powerlaw.

% cut-off-pl Model
\begin{equation}
\begin{split}
    A(E) &= KE^{-\alpha}{\rm exp}(-\frac{E}{E_{\rm cut}}), \\ E_{\rm peak} &= (2-\alpha)E_{\rm cut},
\end{split}
\label{equ:Cutoffpl_Model}
\end{equation}
where $K$ is the normalization amplitude constant ($\rm photons \cdot cm^{-2} \cdot s^{-1} \cdot keV^{-1}$) at 1 keV, $\alpha$ is the power law photon index, and $E_{\rm cut}$ is the e-folding energy of exponential roll-off in keV.

\begin{small}
\begin{equation}
    A(E)= \begin{cases} 
    K \bigg (\frac{E}{100}\bigg )^{\alpha} {\rm exp}\bigg(-\frac{E}{E_{\rm c}}\bigg ), & (\alpha -\beta)E_{\rm c} \geq E, \\ 
    K \bigg [\frac{(\alpha-\beta) E_{\rm c}} {E_{\rm piv}}\bigg ]^{\alpha-\beta} {\rm exp}(\beta-\alpha)\bigg (\frac{E}{100}\bigg )^{\beta}, & (\alpha -\beta)E_{\rm c} \leq E, \\
    \end{cases}
    \label{equ:band_Model}
\end{equation}
\end{small}

where $K$ is the normalization amplitude constant ($\rm photons \cdot cm^{-2} \cdot s^{-1} \cdot keV^{-1}$), $\alpha$ is the first power law photon index, $\beta$ is the second power law photon index, and $E_{\rm c}$ is characteristic energy in keV.

\begin{figure}[http]
\centering
\includegraphics[width=\columnwidth]{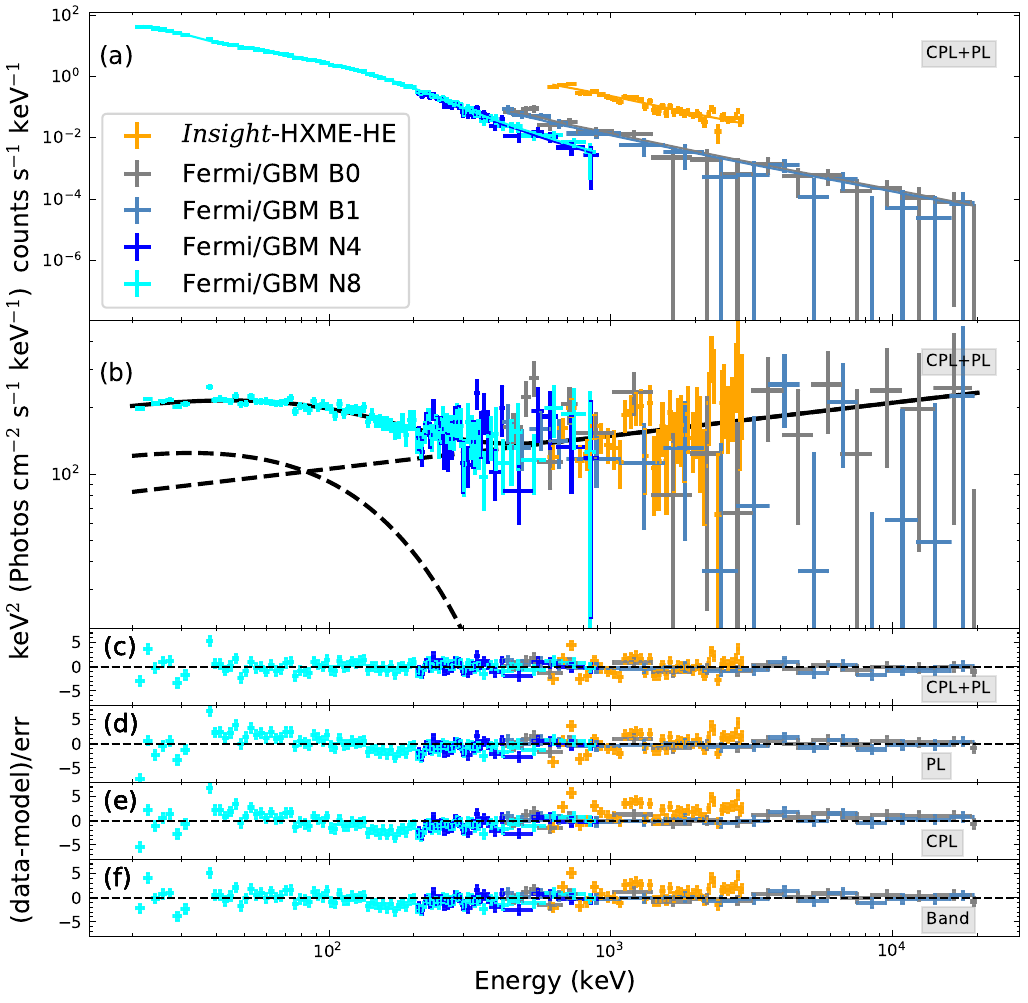}
\caption{\textbf{Panel a}: The count rate spectra in 20-20000 keV energy band from $T_0$+700 s to $T_0$+900 s. \textbf{Panel b}: The spectral energy distribution ($v$F$_{v}$) of CPL+PL model. \textbf{Panel c$\sim$ f}: The residual distribution of CPL+PL, PL, CPL, and Band model.}
\label{spectra_fitting}
\end{figure}

The package of Xspec (V12.12.1) \citep{arnaud1996xspec} is utilized to perform the analysis of those spectra. The X/$\gamma$ spectra in 20-20000 keV energy band are fitted with the PL, CPL, and Band models respectively. As can be seen from Fig. \ref{spectra_fitting}, from $T_0$+700 s to $T_0$+900 s, the spectra fitting results show a significant structure in residual distribution and excess in the low energy band (see Fig. \ref{spectra_fitting} panel d, e, and f). Consequently, we adopt a combined model (CPL+PL) to fit spectra. The combined model fitting results show much better results without significant residual structure and the lowest stat/d.o.f value. Furthermore, the Bayesian information criterion (BIC hereafter), defined as BIC=-2lnL+klnN, is employed to compare and select models \footnote{As mentioned by \cite{delaCruz-Dombriz:2016bqh}, the strength of the evidence supporting the model with the lower BIC value can be summarized as follows: (1) if -2 \textless $\Delta$BIC, no evidence supports the model with a lower BIC value; (2) if -6 \textless $\Delta$BIC \textless -2, positive evidence supports the model with a lower BIC; (3) if -10 \textless $\Delta$BIC  \textless -6, strong evidence supports the model with a lower BIC value; (4) if $\Delta$BIC \textless -10, very strong evidence supports the model with a lower BIC value.}, where L is the likelihood function, k is the number of free parameters of the model, and N is the number of data points \cite[]{2012The}. From $T_0$+660 s to $T_0$+1800 s, the spectra fitting results are summarized in Table \ref{afterglow_table}. The cutoff power-law $E_{\rm peak}$ ($\sim$ 30 keV) seems to be relatively stable during this time range. The power-law model photon indices are also consistent with \insight/HE results ($\sim$ 1.7 from $T_0$+660 s to $T_0$+900 s) within error \citep{HXMT-GECAM}. 

\begin{figure}[http]
\centering
\includegraphics[width=\columnwidth]{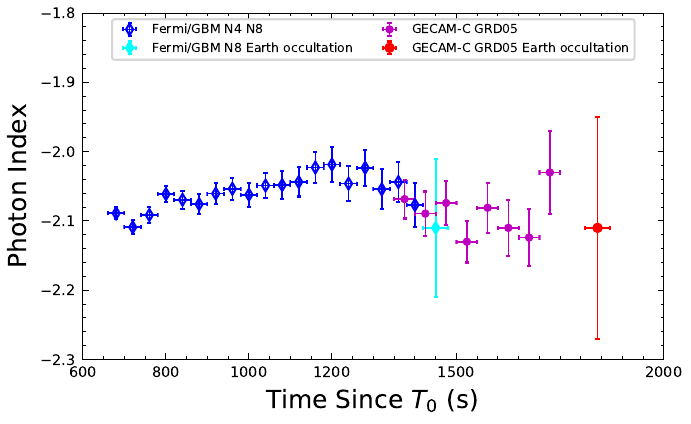}
\caption{The powerlaw photon indices of the early afterglow from $T_0$+660 s to $T_0$+1870 s are derived from \textit{Fermi}/GBM N8 and GECAM-C GRD05 in 20-300 keV energy range. It shows that the photon index first experienced hardening before the break and then softening after the break.}
\label{spectra_evolution}
\end{figure}

Since the source intensity decays with time and the effective area of the detector is minor, in order to guarantee sufficient statistics to obtain meaningful spectra, the detailed analysis of time-resolved spectra is only carried out in the energy range from 20 keV to 300 keV. We generate 19 time-resolved spectra using \textit{Fermi}/GBM N4, and N8 detector data from $T_0$+660 s to $T_0$+1420 s with the same time bin-width of 40 s, and 8 time-resolved spectra using GECAM-C GRD05 detector data from $T_0$+1350 s to $T_0$+1750 s with the same time bin-width of 50 s. The spectra data from 20 keV to 300 keV could be described well with a single power-law model. The photon indices are shown in Fig. \ref{spectra_evolution}. The blue diamonds and magenta points represent the \textit{Fermi}/GBM (N4 and N8) spectra results and GECAM-C (GRD05) spectra results, respectively.

Besides, the source spectrum can also be measured by Earth occultation analysis \citep{Xue:2022bow}. For \textit{Fermi}/GBM, the Earth occultation started at $\sim$ $T_0$+1430 s and ended at $\sim$ $T_0$+1467 s, while it started at $\sim$ $T_0$+1810 s and ended at $\sim$ $T_0$+1860 s for GECAM-C. According to Earth occultation analysis, the photon index of power-law from \textit{Fermi}/GBM is plot with a cyan diamond in Fig. \ref{spectra_evolution}, and that of GECAM-C with a red dot. We note that the photon index measured by these two methods and two instruments is well consistent with each other, demonstrating that the background model used in this work is reasonably reliable.

From the evolution of the photon index in the energy range of 20 keV to 300 keV in Fig. \ref{spectra_evolution}, it seems that the early afterglow first experienced a hardening and then a softening in the spectrum.

\begin{table*}[htbp]
\caption{\centering Fitting time-resolved spectra during the time periods of GRB 221009A early afterglow.}
\begin{tabular*}{\hsize}{@{}@{\extracolsep{\fill}}ccccccc@{}}
\hline
Time$^a$ (s) & Telescopes & CPL $\alpha$  & CPL $E_{\rm cut}$ (keV) & PL $\alpha$ & stat/d.o.f & $\Delta$BIC$^b$\\
\hline
660-700 & \textit{Fermi}/GBM, HXMT/HE  & 1.85$^{+0.05}_{-0.05}$  & 191$^{+54}_{-40}$   & 1.71$^{+0.08}_{-0.10}$ & 375.88/398 & -58.78\\
700-900 & \textit{Fermi}/GBM, HXMT/HE  & 1.69$^{+0.07}_{-0.12}$  & 104$^{+21}_{-22}$   & 1.85$^{+0.05}_{-0.06}$ & 495.51/398 & -174.85\\
900-1400 & \textit{Fermi}/GBM   & 1.52$^{+0.16}_{-0.38}$  & 87$^{+29}_{-29}$   & 1.90$^{+0.06}_{-0.07}$ & 351.81/351 & -80.63\\
1350-1800 & GECAM-C  & 1.30$^{+0.38}_{-0.55}$   & 43$^{+24}_{-15}$  & 1.76$^{+0.19}_{-0.18}$ & 143.11/144 & -9.01\\
\hline
\end{tabular*}
\footnotesize{$^a$ The Time is relative to $T_{\rm AG}$. \\ $^b$ $\Delta$BIC = BIC$_{\rm CPL+PL}$ - BIC$_{\rm PL}$}\\
\label{afterglow_table}
\end{table*}

\subsection{Flux light curves analysis}
\begin{figure}[http]
\centering
\includegraphics[width=\columnwidth]{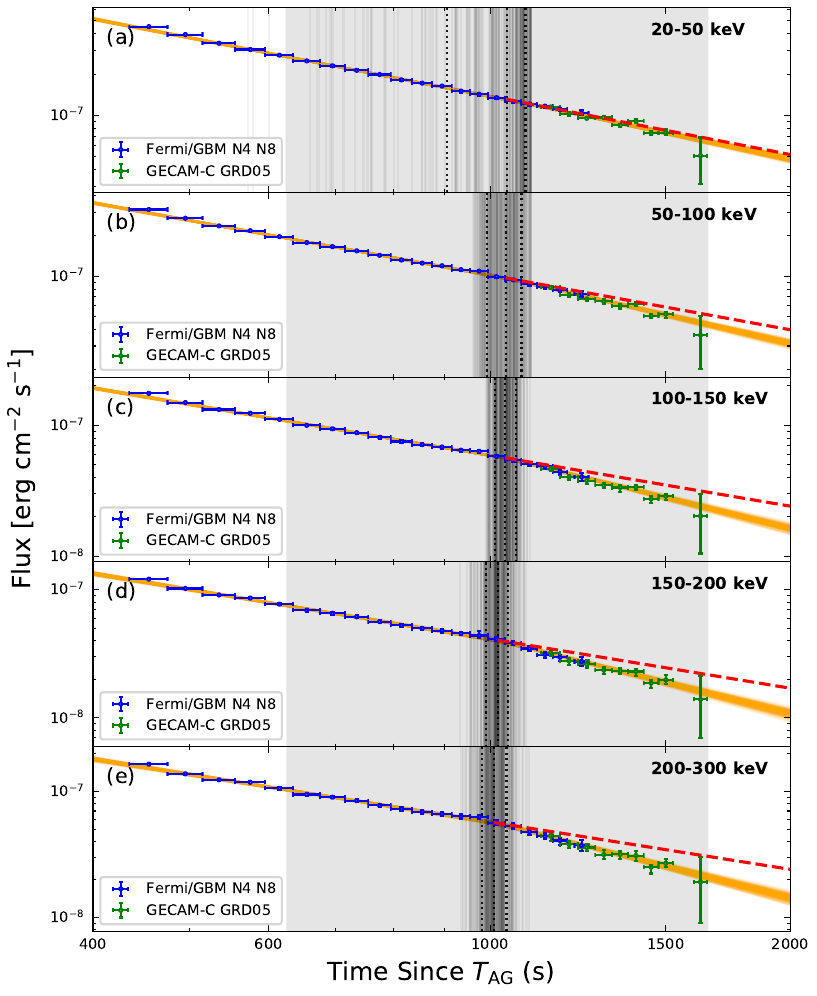}
\caption{The early afterglow flux evolution from $T\rm_{AG}$+435 s to $T\rm_{AG}$+1635 s are given by \textit{Fermi}/GBM and GECAM-C in five energy bands, namely 20-50 keV, 50-100 keV, 100-150 keV, 150-200keV, and 200-300 keV from panel $(a)$ to panel $(e)$. The gray shaded area represents the fitting range. The orange lines show the posterior distribution of BPL, and the gray vertical lines show the posterior distribution of break time. The red dotted lines show the flux decay with the pre-break slope.}
\label{flux_lightcurve}
\end{figure}

According to the results of the spectral analysis conducted with time resolution, the flux light curves (in units of erg/cm$^2$/s) are calculated in five energy bands: 20-50 keV, 50-100 keV, 100-150 keV, 150-200 keV, 200-300 keV, respectively. With the starting time of the afterglow at $T\rm_{AG}$ = $T_0$ + 225 s, as shown in Fig. \ref{flux_lightcurve}, the blue points represent \textit{Fermi}/GBM flux and the green points represent GECAM-C flux in each panel. It is worth noting that the measured flux from both instruments is highly consistent with each other during the overlapping time period, demonstrating that the calibration and background model used in this work are reasonably reliable for \textit{Fermi}/GBM and GECAM-C.

The flux light curves from $T\rm_{AG}$+525 to $T\rm_{AG}$+1635 s are fitted by the single power-law model (Equ.\ref{flux_PL}) and broken power-law model (Equ.\ref{flux_BPL}) \cite[]{2006Physical} with the $emcee$ software package \cite[]{foreman2013emcee}, respectively. The free parameters of the broken power-law model include amplitude $A_{\rm BPL}$, break time $t_{\rm j}$, and two slope indices $\alpha_1$ and $\alpha_2$. The posterior distributions of the broken power-law model and break time are plotted in different energy bands, respectively (see Fig.\ref{flux_lightcurve}). Detailed fitting results for each energy band are listed in Table.\ref{BPL_pars}.

\begin{equation}
    f_{\rm PL}(t) = A_{\rm PL} t ^ {-\alpha}.
\label{flux_PL}
\end{equation}

\begin{equation}
\begin{split}
        f_{\rm BPL}(t) = \left \{
         \begin{array}{ll}
           A_{\rm BPL} (t / t_{\rm j}) ^ {-\alpha_1}, & t < t_{\rm j} \\
           A_{\rm BPL} (t / t_{\rm j}) ^ {-\alpha_2}, &  t > t_{\rm j} \\
         \end{array}
       \right.
\end{split}
\label{flux_BPL}
\end{equation}

\begin{figure*}
\centering
\begin{minipage}[t]{0.54\textwidth}
\centering
\includegraphics[width=\columnwidth]{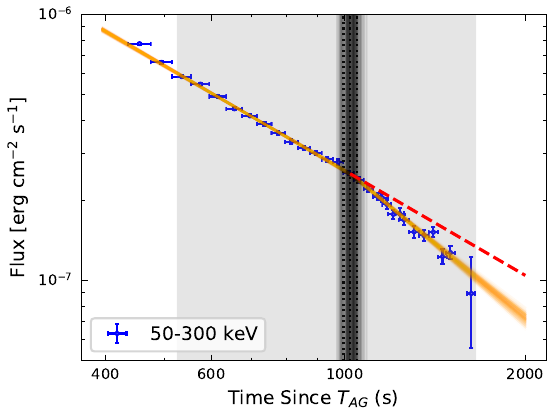}
\end{minipage}
\begin{minipage}[t]{0.40\textwidth}
\centering
\includegraphics[width=\columnwidth]{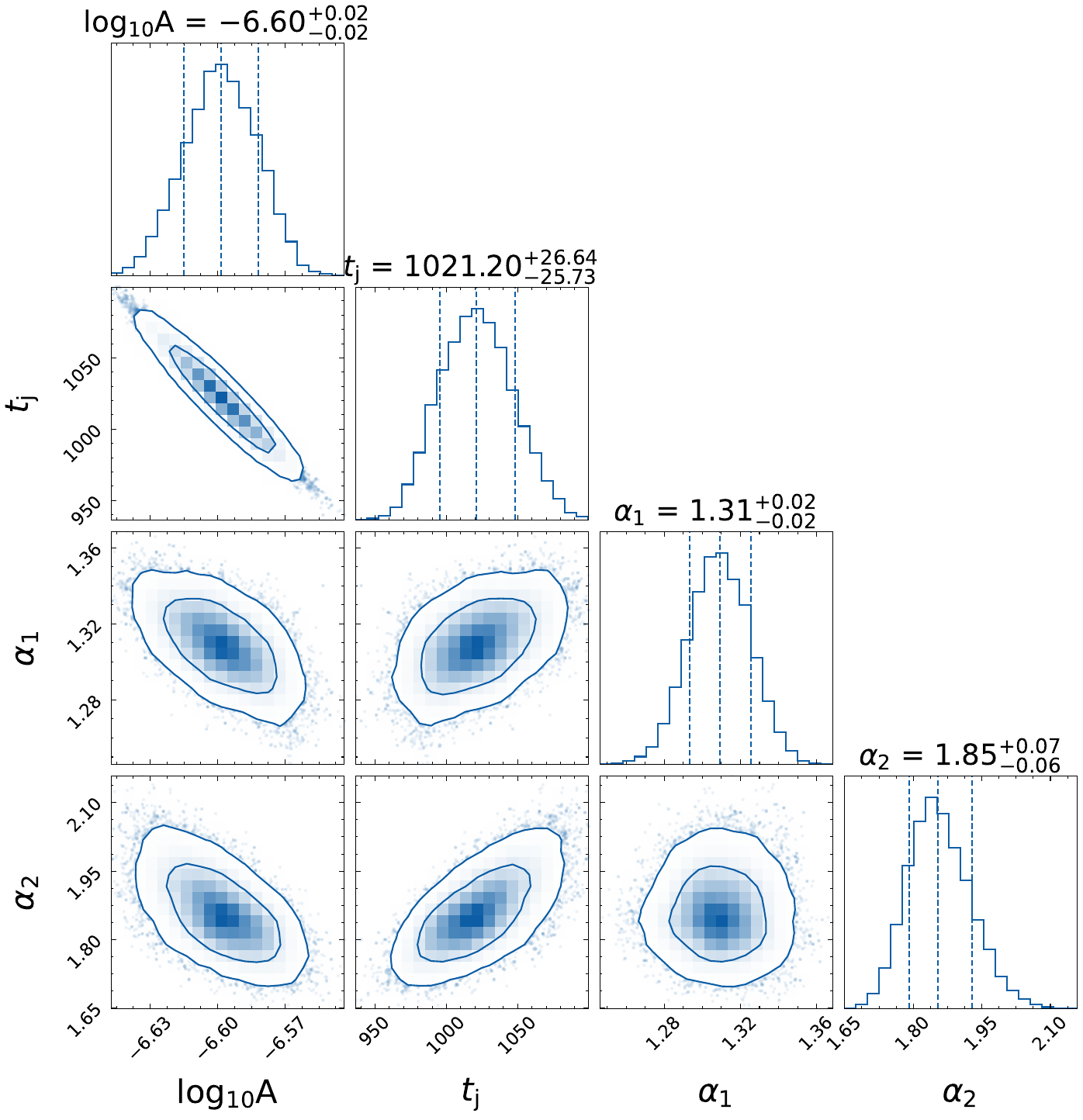}
\end{minipage}
\caption{\textbf{Left}: The early afterglow flux light curve (50-300 keV) jointly observed by GECAM-C and \textit{Fermi}/GBM. The gray shaded area represents the fitting range. The orange lines show the posterior distribution of BPL model, and the gray vertical lines show the posterior distribution of break time. The red dotted lines show the flux decay with the pre-break slope. \textbf{Right}: The corner plot represents the BPL fitting result of the flux light curve in the left panel.}
\label{flux_50-300keV}
\end{figure*}

\begin{figure}[http]
\centering
\includegraphics[width=\columnwidth]{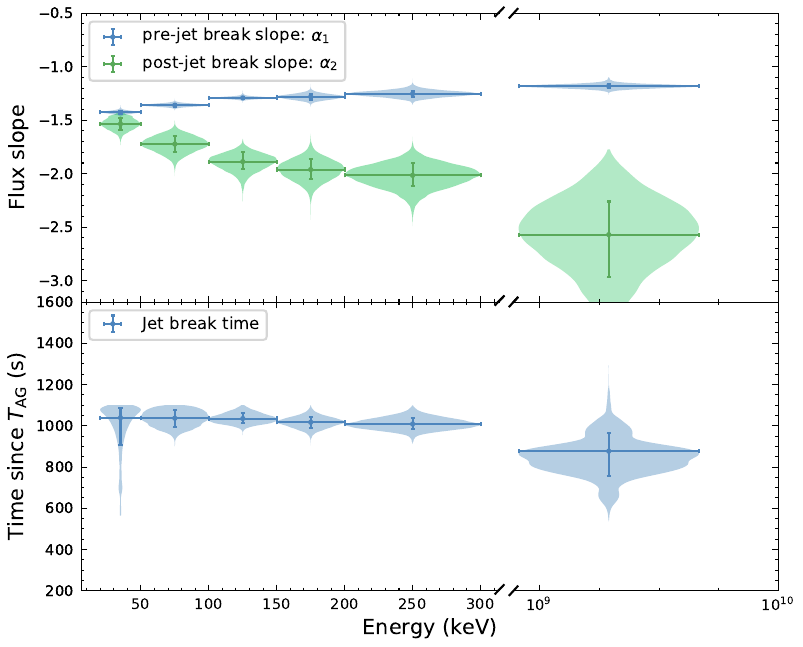}
\caption{\textbf{Top}: Light curve decay slopes before and after the jet break in different energy bands, with error bars indicating 1 $\sigma$ uncertainty. \textbf{Bottom}: The break time of early afterglow in different energy bands, with error bars indicating 1 $\sigma$ uncertainty. The results in the keV energy band are from joint analysis of GECAM-C and \textit{Fermi}/GBM. Our fitting resutls of the jet break in the TeV energy band is consistent with LHAASO team \citep{science.adg9328} within 3 $\sigma$ uncertainty. Notably, the break time seems to be constant from keV to TeV within the error bar.}
\label{figBPL_pars}
\end{figure}

The significance of the break was estimated by the likelihood ratio (LR) test method \citep[e.g.][]{science.adg9328}. All LR significances for $>$ 50 keV energy bands are higher than 3$\sigma$  (50-100 keV: 3.8$\sigma$, 100-150 keV: 6.4$\sigma$, 150-200 keV: 4.4$\sigma$, and 200-300 keV: 4.5$\sigma$), demonstrating that a break is required significantly. Although the LR significance in 20-50 keV energy band is only 1.9 $\sigma$, the $\chi^2$/d.o.f value of the BPL model (1.15) is clearly preferred over the PL model (1.37). We note that this lower significance in low energy bands than higher energy bands seems to be understandable and expected from the trend of the energy dependence of the pre-break and post-break slopes (Fig.\ref{figBPL_pars}). Therefore, we conclude that the BPL model is statistically more suitable to describe the flux decay. We also plot the summed flux light curves in the 50-300 keV and fit with the BPL model from $T\rm_{AG}$+525 to $T\rm_{AG}$+1635 s. The MCMC fit results are shown in Fig.\ref{flux_50-300keV}.

The pre-break, post-break, and break time parameters in the keV-MeV and TeV energy band are shown in Fig.\ref{figBPL_pars} with error bars of 1 $\sigma$ uncertainty. Interestingly, the break time is well-consistent among all energy bands. Fitting the light curve in 50-300 keV yields the break time $t_{\rm j}$ = $T_{AG}$ + 1021$^{+27}_{-26}$ s. We note that this break time derived in the present work is consistent with the results given by the joint analysis of \insight/HE and GECAM-C ($t_{\rm j} \sim T_{\rm AG}+950_{-50}^{+60}$ s in 20-200 keV) \citep{HXMT-GECAM}.  Our result is also supported by a recent independent study with \textit{Fermi}/GBM data only \citep{Zhang:2023dnq}. 
Furthermore, we also fit the flux light curve in TeV energy band from $T_{AG}$ + 100 s to $T_{AG}$ + 3000 s (ignored time range of $T_{AG}+[255, 525]$ s to avoid potential contamination from the flare) using BPL model. The flux decay slopes and break time are consistent with \cite{science.adg9328} within 3 $\sigma$ uncertainty, as shown in Fig.\ref{figBPL_pars}.

We note that the break time seems to have a marginal dependence on the energy band. We testify this possible dependence by fitting the break time as a function of energy with the PL model and constant model respectively. The $\Delta BIC$ ($BIC_{PL}$-$BIC_{const}$) of these two models is only -1.1, meaning that there is no statistical evidence for the energy dependence of the break time. Consequently, we conclude that the break time remains constant in the very wide energy band from keV to TeV, provides compelling evidence for the origin of jet break. Interestingly, we find that the light curve decay slope varies with energy from keV to TeV in an opposite way: the pre-jet break slope $\alpha_{1}$ increases with energy, while the post-jet break slope $\alpha_{2}$ decreases with energy, as shown in Fig.\ref{figBPL_pars}.

\begin{table*}[htbp]
\caption{\centering Fitting results of the flux light curves during the periods of early afterglow.}
\begin{tabular*}{\hsize}{@{}@{\extracolsep{\fill}}ccccccccccccc@{}}
\hline

\multirow{2}{*}{Energy band (keV)} & \multicolumn{5}{c}{BPL} & \multicolumn{3}{c}{PL} & \multirow{2}{*}{$\Delta$BIC}$\dagger$ \\
\cmidrule(r){2-6} \cmidrule(r){7-9}
 & slope ($\alpha_1$) & break time ($t_{\rm j}$) $^*$ (s) & slope ($\alpha_2$) & $\chi^2$/d.o.f & BIC$\rm _{BPL}$ & slope($\alpha$) & $\chi^2$/d.o.f &  BIC$\rm _{PL}$ &\\
 
\hline
20-50 & 1.43$^{+0.01}_{-0.01}$ & 1047$^{+41}_{-111}$ & 1.54$^{+0.06}_{-0.05}$  & 26.47/23 & 39.65 & 1.45$^{+0.01}_{-0.01}$ & 34.29/25 & 40.88 & -1.23 \\
50-100 & 1.36$^{+0.01}_{-0.01}$ & 1036$^{+37}_{-41}$ & 1.72$^{+0.07}_{-0.07}$ & 23.84/23 & 37.02 & 1.41$^{+0.01}_{-0.01}$ & 76.99/25 & 83.58 & -46.56\\
100-150 & 1.29$^{+0.01}_{-0.01}$ & 1035$^{+27}_{-23}$ & 1.89$^{+0.09}_{-0.07}$ & 28.48/23 & 41.66 & 1.32$^{+0.01}_{-0.01}$ & 190.81/25 & 197.40 & -155.74\\
150-200 & 1.28$^{+0.02}_{-0.03}$ & 1018$^{+25}_{-25}$ & 1.96$^{+0.10}_{-0.09}$ & 23.24/23 & 36.42 & 1.41$^{+0.02}_{-0.01}$ & 76.75/25 & 83.34 & -46.92\\
200-300 & 1.25$^{+0.03}_{-0.03}$ & 1008$^{+30}_{-26}$ & 2.02$^{+0.11}_{-0.10}$ & 16.63/23 & 29.81 & 1.40$^{+0.02}_{-0.02}$ & 71.22/25 & 77.81 & -48.00\\
\hline
50-300 & 1.31$^{+0.02}_{-0.02}$ & 1021$^{+27}_{-26}$ & 1.85$^{+0.07}_{-0.06}$ & 19.79/23 & 32.97 & 1.40$^{+0.01}_{-0.01}$ & 66.03/25 & 72.62 & -39.65\\
\hline

\end{tabular*}
\footnotesize{$^*$ The break time ($t_{\rm j}$) is relative to afterglow starting time $T \rm _{AG}$ = $T_0$+225 s. \\ $\dagger$ $\Delta$BIC= BIC$\rm _{BPL}$ - BIC$\rm _{PL}$.}\\
\label{BPL_pars}
\end{table*}

Following the calculation in \cite{HXMT-GECAM}, the jet opening angle can be inferred with \citep{2018pgrb.book.....Z}:
\begin{equation}
    \begin{aligned}
        \theta_{\rm j} &\simeq (0.063\, {\rm rad})\left({\frac{t_{\rm j}}{1\ {\rm day}}}\right)^{\frac{3}{8}} \left(\frac{1+z}{2}\right)^{-\frac{3}{8}} \\ & \left(\frac{{E_{\gamma,{\rm iso}}}}{10^{53}\ {\rm erg}}\right)^{-\frac{1}{8}} \times \left(\frac{\tilde{{\eta}_{\gamma}}}{0.2}\right)^{\frac{1}{8}} \left(\frac{n}{0.1\ {\rm cm}^{-3}}\right)^{\frac{1}{8}}.
    \end{aligned}
\label{openangle}
\end{equation}
Considering the measured jet break time $t_{\rm j}$, $E_{\gamma,{\rm iso}}$ and redshift $z$, we derive a jet opening angle of $\theta_{\rm j} \sim {0.73^\circ}^{+0.01^\circ}_{-0.01^\circ} (\tilde\eta_\gamma n)^{1/8}$, where $\tilde{\eta_\gamma}$ is the ratio between isotropic $\gamma$-ray and isotropic kinetic energies, and $n$ is the ambient number density in units of $\rm cm^{-3}$, which are left as free parameters.

\section{Discussion and Conclusion} \label{section5}
GRB 221009A is the Brightest Of All Time (BOAT) gamma-ray burst ever detected. In this letter, we report a comprehensive temporal and spectral analysis on its early afterglow in the keV-MeV energy band (20 keV - 20 MeV) from $T_0$+660 s to $T_0$+1860 s with combined observation data from \textit{Insight}-HXMT, GECAM-C and \textit{Fermi}/GBM. 

We find that the spectrum of early afterglow in 20 keV - 20 MeV can be well described with the CPL+PL model, where the CPL and PL dominate the low-energy and high-energy bands respectively. Both the CPL $E_{\rm peak}$ ($\sim$30 keV) and the PL photon index ($\sim$1.8) seem generally stable from $T_0$+660 s to $T_0$+1860 s. We also note that the CPL $E\rm _{peak}$ and low energy photon indices are broadly consistent with the later afterglow results ($E\rm _{break}$=33.6$^{+16.1}_{-28.6}$ keV, $\Gamma$=1.7 @ $T_0$+4.2 ks) within the error \cite[Table.1 \& Table.2]{williams2023grb}. Due to the relatively large error on $E\rm _{peak}$, it is unclear whether and how the $E\rm _{peak}$ evolves with time. Furthermore, the high energy PL component may also be detected by \textit{Swift}/BAT in later afterglow, because the spectrum residual ratio shows clear evidence for the existence of an excess in \textit{Swift}/BAT energy band \cite[the right panel of Fig.5]{williams2023grb}.

This CPL+PL spectrum shape is very unusual in GRB afterglow since the synchrotron radiation mechanism usually gives rise to a Band shape or power-law shape spectrum. But the GRB 221009A early afterglow shows a "V" shape spectrum above 20 keV. Indeed, this kind of spectrum shape was also reported in GRB 990123 afterglow \citep{2005The_Corsi, 2005The_Maiorano}. However, the high-energy spectrum from GRB 990123 afterglow was only detected up to 60 keV and at a relatively later time (from $\sim$20 min to 60 min after the GRB trigger), while the early afterglow of GRB 221009A was measured in an extremely wide band (including keV, MeV, GeV, and TeV energy band) and at a very early stage (immediately after the prompt emission and flare episode) \citep[e.g.][]{HXMT-GECAM, GBM_09A, science.adg9328, Zhang:2023dnq}. Further promoted by the exceptional brightness of this GRB, the detailed and accurate measurement of the early afterglow spectrum of GRB 221009A is truly unprecedented in GRB observation history. The origin of this early afterglow spectrum is intriguing and subject to further studies.

For the refined spectral analysis with a shorter time bin, a single PL model can describe well in 20-300 keV energy band, based on which we estimate the flux light curves in five energy bands. The PL photon index ($\sim$ -2.1) is in good agreement with \textit{Swift}/BAT results ($\Gamma$=-2.08$_{+0.03}^{-0.03}$ @ $T_0$+3302 $\sim$ $T_0$+4538 s)\citep{williams2023grb}.

Although the wide band (from 20 keV to 20 MeV) spectral shape (CPL+PL) of the early afterglow did not change significantly, the spectral hardness experienced a pre-break hardening and post-break softening (Fig. \ref{spectra_evolution}), which is somewhat consistent with the spectral evolution in the TeV energy band \citep[Fig. 3]{science.adg9328}. 

A significant and achromatic break is found at $t_{\rm j}$ = $T_{AG}$ + 1021$^{+27}_{-26}$ s in the 50-300 keV light curve of early afterglow, 
supporting that this is a jet break. Remarkably, this jet break time in 50-300 keV is well consistent with that in TeV energy band \citep{science.adg9328}. More interestingly, we find that the pre-break and post-break slopes of the flux light curve vary with energy (from keV to TeV) clearly. These two slopes become closer as energy decreases, making the break less identifiable in the low energy band. Remarkably, our results show that the pre-break and post-break slopes tend to converge to a value of about -1.5 in the low energy band (20-50 keV), which is well consistent with the pre-break slope of -1.498$^{+0.004}_{-0.004}$ reported in the soft X-ray energy band (0.3-10 keV) \citep{williams2023grb}. We therefore suggest that this jet break feature may disappear in the lower energy band (i.e. soft X-ray and even lower energy band), thus forming an apparent peculiar break feature. Although the observations of a break in the X-ray but not in the optical energy band have been reported \citep{Panaitescu:2006yj, 2008Broadband}, here for the first time, we reveal the detailed transition from the break feature in higher energy to the non-break in lower energy. On the other hand, the energy-dependent slopes before and after the jet break are not expected from the ideal top-hat jet model \citep{2018pgrb.book.....Z}, thus this observation strongly disfavors the simple top-hat jet model and calls for a more sophisticated jet structure and new physics involved.

We note that, compared to our finding of the jet break at an early time ($T_{\rm AG}$ + $\sim$1021 s), there are several breaks in later time reported in the multi-wavelength observation of GRB 221009A afterglow. For example, a break at 7.9$_{-1.9}^{+1.1} \times 10^4$ s is reported in the soft X-ray light curve\citep{williams2023grb}, while an achromatic break at $\sim$0.6 day is claimed in the optical bands \citep{Shrestha:2023ooz}. While these breaks are measured in a relatively limited energy band, our jet break reported here is seen in a very wide energy band from keV to TeV band. We also note that, for our jet break, the post-break slope in high energy band (200-300 keV) is close to -2, which is expected by the afterglow theory \citep{2006Physical}, while the post-break slope in soft X-ray and optical is much shallower than theory expectation \citep{williams2023grb, Shrestha:2023ooz}.
Moreover, the jet break at a much earlier time indicates a very narrow jet core which is well concordant with the extremely luminous of the GRB.

\section{Summary}
In this work, we have performed an unprecedented and detailed temporal and spectral analysis of the early afterglow (from $T_0$+660 s to $T_0$+1860 s) of the Brightest Of All Time (BOAT) GRB 221009A in X/$\gamma$-ray energy band with joint observation of GECAM-C, \textit{Fermi}/GBM and \insight. We discovered many features in the early afterglow of this exceptional burst, including unusual spectrum (CPL+PL, basically a V-shape from 20 keV to 20 MeV), achromatic jet break time (from keV to TeV) and energy-dependent flux decay slopes before and after the jet break, etc. We point out that these observational features and properties may set a new benchmark for GRBs, thanks to the exceptional brightness and observation coverage of this GRB 221009A. We argue that these results strongly disfavor the simple top-hat jet model but provide support to the structured jet scenario with a bright narrow core. This is also consistent with the statistical analysis results by comparing GRB 221009A with the entire GRB sample \citep{Lan:2023khy}. In this structured jet model, the narrow jet core with ultra-relativistic Lorentz factor should be surrounded by a wider jet with a lower Lorentz factor, leading to complicated temporal and spectral features \citep[e.g.][]{Panaitescu:2006yj, 2008Broadband}. A joint analysis of the early and late afterglow of GRB 221009A could further deepen our understanding of GRB and jet physics. \citep[e.g.][]{Ren:2022icq, Sato:2022kup, OConnor:2023ieu, Gill:2023ijr, Zhang:2023pro}.

\section*{Acknowledgments}
The authors thank the support from the National Key R\&D Program of China (Grant No. 2021YFA0718500), % HXMT & GECAM
the Strategic Priority Research Program of Chinese Academy of Sciences (Grant No. XDA15360102, XDA15360300, XDA15052700), % GECAM
the National Natural Science Foundation of China 
(Grant No. 12273042, % xiong shaolin
12273043, % Li xiaobo
U2038106 %li chengkui
12173038, % li xinqiao
), 
the Science Foundation of Hebei Normal University (No. L2023B11). % cai ce
 
The GECAM (Huairou-1) mission is supported by the Strategic Priority Research Program on Space Science of the Chinese Academy of Sciences.
This work made use of the data from the {\it Insight}-HXMT mission, a project funded by the China National Space Administration (CNSA) and the Chinese Academy of Sciences (CAS).
We appreciate the GECAM and {\it Insight}-HXMT teams.
We acknowledge the public data and software from {\it Fermi}/GBM. 
We appreciate helpful discussions with Zhen Cao, Min Zha, Bing Zhang, Xiangyu Wang, Zigao Dai, Binbin Zhang, Yongfeng Huang, Yunwei Yu, Sarah Antier and Eric Burns. 
We appreciate the valuable comments and suggestions from anonymous referee.

\bibliography{reference}
\bibliographystyle{aasjournal}

\end{document}